\documentclass[a4paper,12pt]{article}

\usepackage{color}
\usepackage[utf8]{inputenc}
\usepackage[english]{babel}
\usepackage{amssymb}
\usepackage{amsmath}
\usepackage{graphicx}
\usepackage{hyperref}
\usepackage{amsfonts}
\usepackage{latexsym}
\usepackage{epsf,epsfig,rotating}
\usepackage{cite}

\usepackage[left=3cm,right=2.5cm,top=2cm,bottom=3cm,bindingoffset=0cm]{geometry}

\begin{document}
\renewcommand\bibname{\Large References}

\title{
\begin{flushright}
{\small INR-TH-2016-036}
\end{flushright}
Q-holes}

\author{
E.\;Nugaev$^{a,b}$\thanks{{\bf e-mail}:
emin@ms2.inr.ac.ru}, A.\;Shkerin$^{c,a}$\thanks{{\bf e-mail}:
shkerin@inr.ru}, M.\;Smolyakov$^{d,a}$\thanks{{\bf e-mail}: smolyakov@theory.sinp.msu.ru}
\\
$^a${\small{\em
Institute for Nuclear Research of the Russian Academy
of Sciences,}}\\
{\small{\em 60th October Anniversary prospect 7a, 117312, Moscow,
Russia
}}\\
$^b${\small{\em Moscow Institute of Physics and Technology,
}}\\
{\small{\em Institutskii per. 9, Dolgoprudny, Moscow Region 141700,
Russia
}}\\
$^c${\small{\em \'{E}cole Polytechnique F\'{e}d\'{e}rale de Lausanne,
}}\\
{\small{\em  CH-1015, Lausanne, Switzerland}}\\
$^d${\small{\em Skobeltsyn Institute of Nuclear Physics, Moscow
State University,
}}\\
{\small{\em 119991, Moscow, Russia}}
}
\date{}

\maketitle

\begin{abstract}
We consider localized soliton-like solutions in the presence of a stable scalar condensate background. By the analogy with classical mechanics, it can be shown that there may exist solutions of the nonlinear equations of motion that describe dips or rises in the spatially-uniform charge distribution. We also present explicit analytical solutions for some of such objects and examine their properties.
\end{abstract}

\section{Introduction}

Spatially-homogeneous solutions in the complex scalar field theories with the global $U(1)$-invariance have been proven to be very useful in different branches of modern physics. Perhaps the most known example of their application to cosmology is the Affleck-Dine mechanism of baryogenesis \cite{Affleck:1984fy}. Evolution of the spatially-homogeneous condensate in the Early Universe, which is usually studied numerically, is subject to certain restrictions in order to yield a successful cosmological scenario \cite{Doddato:2011fz}. For instance, a possible spatial instability of the condensate results in its fragmentation into nontopological solitons --- Q-balls. The latter, in turn, can be a crucial ingredient in the solution of the dark matter problem \cite{Kusenko:1997si}. This makes inhomogeneous classical solutions also of considerable interest in cosmology. Their another application is related to the possibility of production of gravitational waves \cite{Zhou:2015yfa,Antusch:2016con,Katz:2016adq}.

Emergence of localized stationary configurations was first discovered in the systems whose evolution is governed by the Nonlinear Schr\"{o}dinger Equation (NSE) \cite{Zakharov}. In nonlinear optics these solutions are known as bright solitons. Similar solutions in a theory of the complex scalar field in four dimensional space-time, possessing the global $U(1)$-charge, were called ``Q-balls'' by S. Coleman \cite{Coleman:1985ki}. NSE admits another interesting class of solutions corresponding to ``dark solitons'' in a stable medium \cite{Zakharov_Shabat}. They have the form of a dip in a homogeneous background. It is important to note that these solutions are of the topological nature. In particular, they cannot be deformed into the surrounding condensate by a finite amount of energy. Therefore, the question arises about the existence and properties of the analogs of dark solitons in the complex scalar field theory, where they presumably can be analyzed by the same methods as the ordinary Q-balls.

The existence of the dip-in-charge-like solutions in scalar field theories is not a manifestation of some specific properties of these theories. In fact, such solutions exist for the usual ``Mexican hat'' scalar field potential. To see this, let us consider the complex scalar field $\phi$ with the Lagrangian density
\begin{equation}
\partial^{\mu}\phi^*\partial_{\mu}\phi-\frac{\lambda}{2}(\phi^*\phi-v^2)^2.
\label{phi4}
\end{equation}
If $\lambda>0$, the theory admits the well-known real static solution --- the kink, which has the form
\begin{equation}\label{RealKink}
\phi=v\tanh\left(\sqrt{\frac{\lambda}{2}}v x\right).
\end{equation}
It can be generalized to a class of stationary but not static solutions as follows,
\begin{equation}
\phi=e^{i\omega t}f(x),
\label{anz1_1}
\end{equation}
where $\omega$ is a constant parameter and
\begin{equation}\label{chargedkink}
f(x)=\sqrt{v^2+\frac{\omega^2}{\lambda}}\tanh\left(\sqrt{\frac{\lambda}{2}\left(v^2+\frac{\omega^2}{\lambda}\right)}\,x\right).
\end{equation}
Then, for the $U(1)$-charge density $\rho$ we get
\[
\rho=2\omega f^2,
\]
which clearly has a dip around the origin $x=0$. The kink solution (\ref{RealKink}) is unstable in this model and can be interpreted as a sphaleron in the Abelian gauged version of (\ref{phi4}), see \cite{Bochkarev:1987wg} for details. Another distinctive feature of the model (\ref{phi4}) is the stability of the charged condensate as long as $\lambda>0$. We note that the solution (\ref{anz1_1}) requires an infinite amount of energy to be deformed into the spatially-homogeneous condensate of the same charge or frequency.

In this paper we present the soliton-like localized solutions in a theory of the complex scalar field, which describe inhomogeneities in the charge distribution of the condensate and can be deformed into the spatially-homogeneous condensate of the same frequency using a finite amount of energy. We will refer to such solutions as ``Q-holes'' or ``Q-bulges'' in order to stress their similarity to the ordinary Q-balls and to the ``holes in the ghost condensate'' of \cite{Krotov:2004if}. In the next section we will argue in favor of existence of these solitons with the help of Coleman's overshoot-undershoot method and survey their general properties. In Section~\ref{Explicit examples of Q-holes} we will present and examine the explicit examples of Q-holes in one and three spatial dimensions. In Section~4 we will discuss the classical stability (in fact, instability) of Q-holes and Q-bulges. In Conclusion we will briefly discuss the obtained results.

\section{General considerations}
\subsection{Q-balls}
Consider a theory of the complex scalar field $\phi$ in flat $(d+1)$-dimensional space-time, with the action
\begin{equation}\label{action}
S=\int dt d^{d}x\left(\partial^{\mu}\phi^{*}\partial_{\mu}\phi-V(\phi^{*}\phi)\right).
\end{equation}
Suppose that the potential $V(\phi^{*}\phi)$ has a minimum (local or global) at $\phi=0$. Then the theory may admit localized configurations called Q-balls \cite{Coleman:1985ki,Rosen0}. They are solutions to the corresponding equations of motion of the form
\begin{equation}\label{qballans}
\phi(t,\vec x)=e^{i\omega t}f(\vec x),
\end{equation}
where $f(\vec x)$ is a real function such that $\lim\limits_{|\vec x|\to\infty}f(\vec x)\to 0$. When $d>1$, it is assumed that $f(\vec x)=f(r)$, where $r=\sqrt{{\vec x}^2}$, $f(r)>0$ for any $r$, and $\partial_{r}f(r)|_{r=0}=0$.

\begin{figure}[h]
\begin{minipage}{0.49\linewidth}
\center{\includegraphics[width=0.75\linewidth]{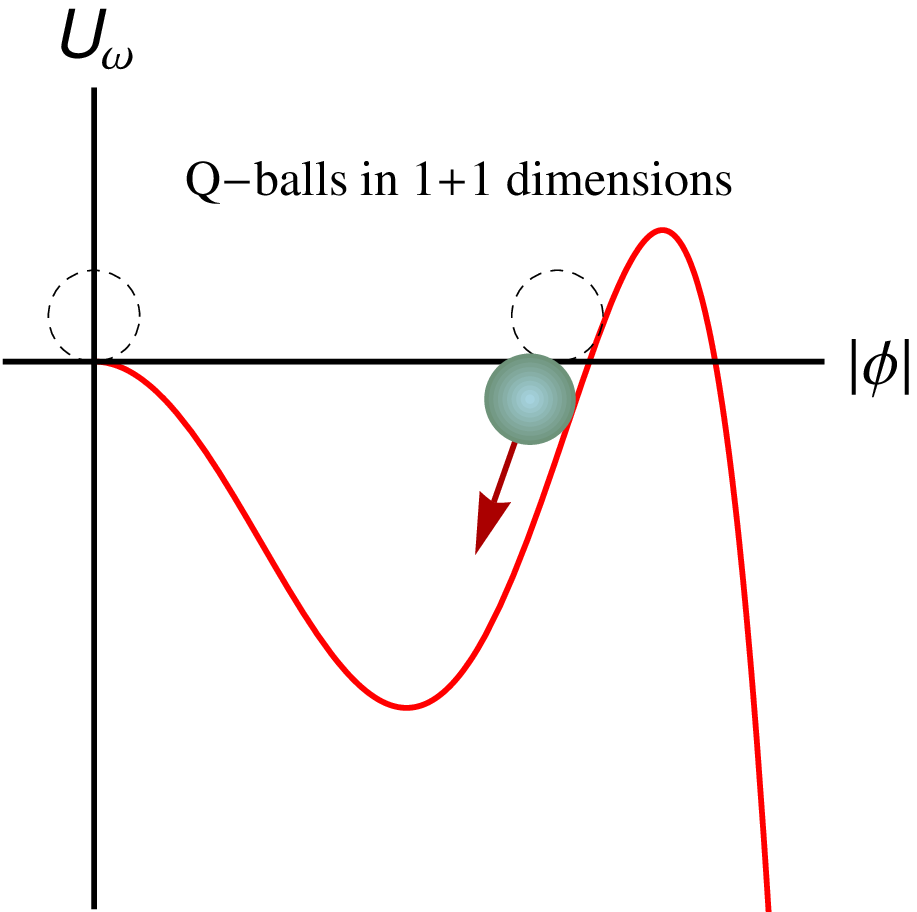}}
\end{minipage}
\begin{minipage}{0.49\linewidth}
\center{\includegraphics[width=0.75\linewidth]{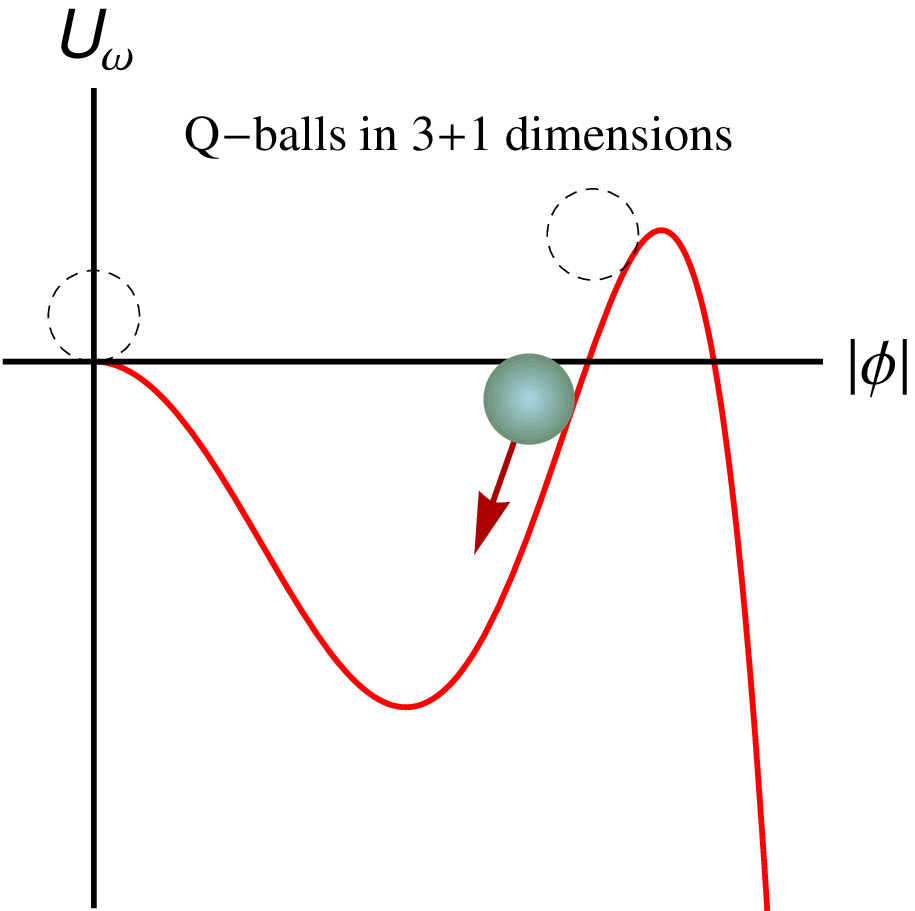}}
\end{minipage}
\caption{Mechanical analogy for Q-balls in $1+1$ and $3+1$ dimensions respectively.}
\label{plotUqballs}
\end{figure}

With the ansatz (\ref{qballans}), the equations of motion for the field $\phi$ reduce to the equation for the function $f$,
\begin{equation}\label{3+1 EoM}
\frac{d^{2}f}{dr^{2}}+\frac{d-1}{r}\frac{df}{dr}+\omega^{2}f-\frac{1}{2}\frac{dV(f)}{df}=0.
\end{equation}
It is a well-known observation that the latter equation can be thought of as an equation of motion of a point particle in classical mechanics, with the ``coordinate'' $f$ and the ``time'' $x$ (or $r$), that moves in the effective potential
\begin{equation}\label{Upotdef}
U_{\omega}(f)=\frac{1}{2}\left(\omega^{2}f^{2}-V(f)\right).
\end{equation}
For $d>1$, the motion of the particle is also affected by the ``friction'' term $\sim\frac{1}{r}\frac{df}{dr}$. This mechanical analogy is illustrated in Fig.~\ref{plotUqballs}.
Specifically, the particle begins to move at the ``moment of time'' $x=0$ (or $r=0$) from the ``coordinate'' $f=f_{max}$ and reaches the vacuum state $f=0$ at the ``time'' $x\to\infty$ (or $r\to\infty$).\footnote{In the one-dimensional case this analogy determines the Q-ball solution only at $x\ge 0$. The full solution is symmetric: $f(-x)=f(x)$.} Note that for $d>1$, $U_{\omega}(f(0))>U_{\omega}(0)$ because of the ``friction'' term.

The reasoning outlined above, despite being simple, can help to unveil a new class of solutions in the case when the potential $U_{\omega}(f)$ possesses other local (or global) maxima except that at $f=0$. In the rest of the paper we will explore this case in detail.

\subsection{Time-dependent scalar condensate, Q-holes and Q-bulges}\label{Time-dependent scalar condensate and Q-holes}
Suppose that for certain values of $\omega$, the effective potential $U_{\omega}(f)$ develops a maximum point,
\begin{equation}\label{fcdef}
\frac{dU_{\omega}(f)}{df}\biggl|_{f=f_{c}}=0,
\end{equation}
at some constant $f_{c}\neq 0$.\footnote{In general, the constants $f_{c}$ are frequency-dependent. In special cases, however, they may be independent of $\omega$.} Then a family of spatially-homogeneous time-dependent solutions appears in addition to the vacuum solution $f\equiv 0$,
\begin{equation}\label{scalcondans}
\phi(t,\vec x)=f_{c}e^{i\omega t},\;\;\;f_{c}\neq 0.
\end{equation}
Without loss of generality, we take $f_{c}$ to be real such that $f_{c}>0$. The solutions (\ref{scalcondans}) represent the scalar condensate and have an infinite total charge and energy. As will be shown below, they can be stable under small fluctuations. Note that, in general, the existence of extra maxima of the effective potential $U_{\omega}(f)$ does not imply the existence of extra minima of the initial potential $V(f)$.

\begin{figure}[h]
\begin{minipage}{0.49\linewidth}
\center{\includegraphics[width=0.75\linewidth]{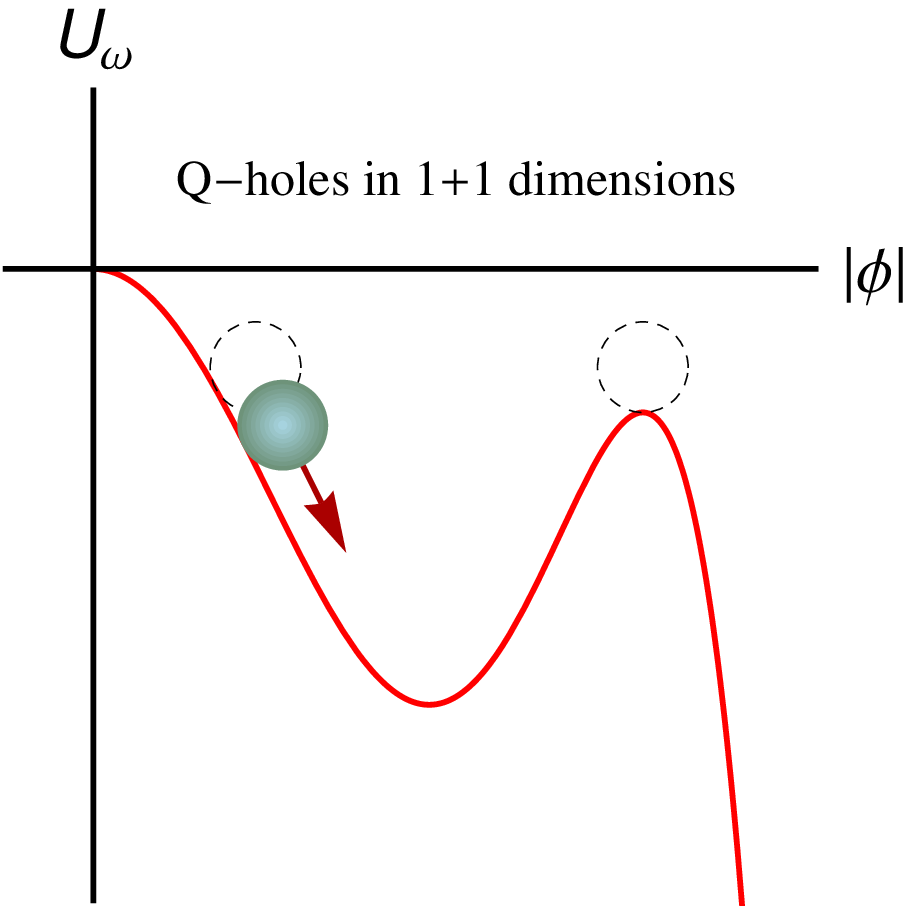}}
\end{minipage}
\begin{minipage}{0.49\linewidth}
\center{\includegraphics[width=0.75\linewidth]{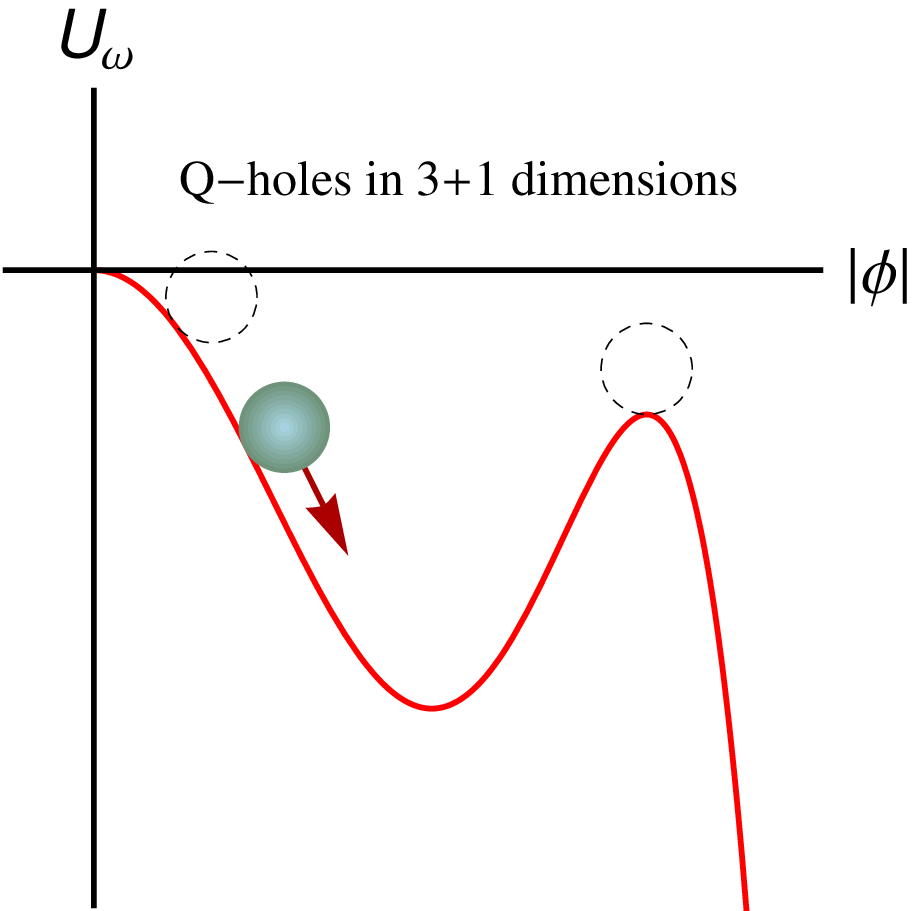}}
\end{minipage}
\caption{Mechanical analogy for Q-holes in $1+1$ and $3+1$ dimensions respectively.}
\label{plotUqholes}
\end{figure}

We can now use the mechanical analogy described in the previous section to advocate the existence of inhomogeneous solutions of the form $\phi(t,\vec x)=f(r)e^{i\omega t}$ in addition to the time-dependent scalar condensate (\ref{scalcondans}).\footnote{Again, without loss of generality, we take $f(r)$ to be real and such that $f(r)>0$.} Here we will discuss two types of such solutions. The mechanical analogy for the first type is presented in Fig.~\ref{plotUqholes}. The crucial feature of these solutions, which we will refer to as ``Q-holes'', is expressed by the inequality $f(\infty)>f(0)$. That is, they can be thought of as ``dips'' in the homogeneous charged condensate.

\begin{figure}[ht]
\begin{minipage}{0.49\linewidth}
\center{\includegraphics[width=0.75\linewidth]{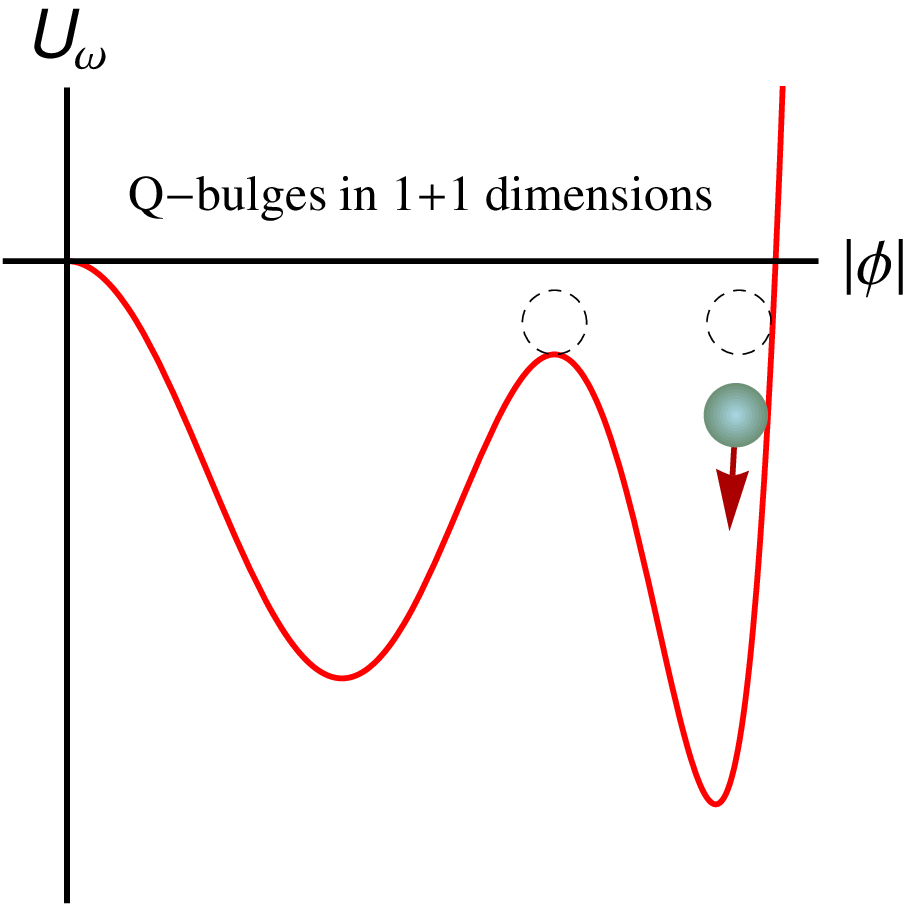}}
\end{minipage}
\begin{minipage}{0.49\linewidth}
\center{\includegraphics[width=0.75\linewidth]{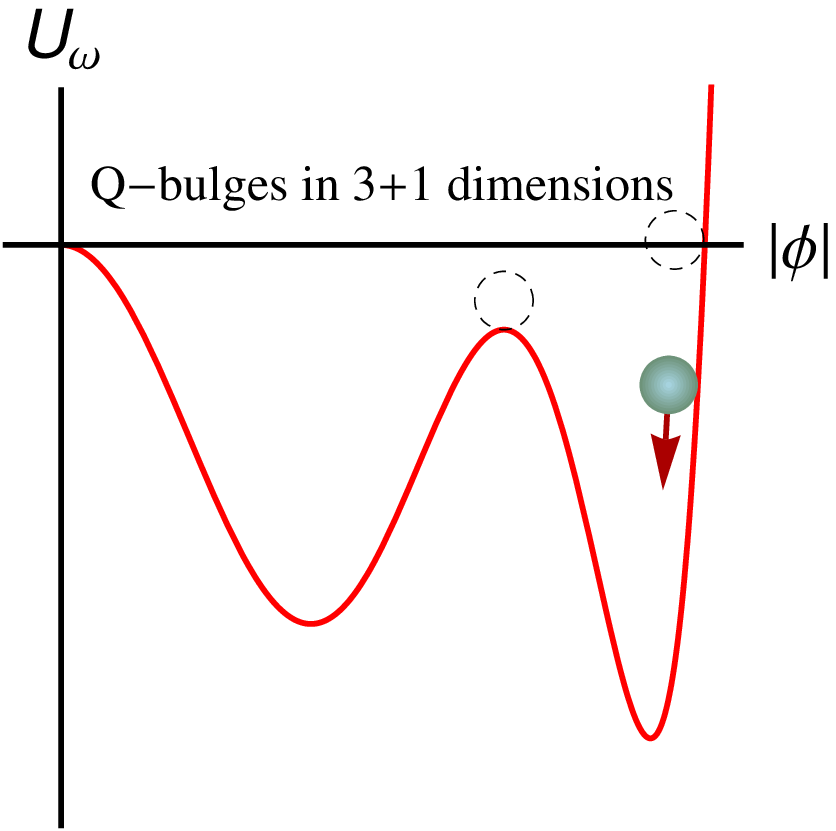}}
\end{minipage}
\caption{Mechanical analogy for Q-bulges in $1+1$ and $3+1$ dimensions respectively.}
\label{plotUbulges}
\end{figure}

The mechanical analogy for the second type is presented in Fig.~\ref{plotUbulges}. Solutions of this type obey the inequality $f(\infty)<f(0)$. Hence they can be thought of as ``rises'' in the homogeneous charged condensate. For this reason, we will call such solutions ``Q-bulges''.

We can see from Fig.~\ref{plotUbulges} that the existence of Q-bulges demands a specific high energy behavior of the effective potential. Apart from this fact, from the point of view of the mechanical analogy, Q-bulges lie close to Q-balls.

Let us briefly discuss the main properties of Q-holes and Q-bulges. First, their asymptotes at infinity,
\begin{equation}
f(\vec x)\to f_{c},\;\;\;|\vec x|\to\infty,
\end{equation}
imply that the frequency $\omega$ of the Q-hole (Q-bulge) is fixed by the frequency of the scalar condensate of magnitude $f_{c}$. Second, the charge and the energy of the Q-hole (Q-bulge) are defined in the standard way,
\begin{eqnarray}\label{Q,E_1}
\begin{aligned}
Q=i\displaystyle\int\left(\phi\partial_{0}\phi^*-\phi^*\partial_{0}\phi\right)d^{d}x=2\omega\displaystyle\int f^{2}d^{d}x,\\
E=\displaystyle\int\left(\omega^{2}f^{2}+\sum\limits_{i=1}^{d}\partial_{i}f\partial_{i}f+V(f)\right)d^{d}x.
\end{aligned}
\end{eqnarray}
When being calculated at a given Q-hole (Q-bulge) configuration, the expressions (\ref{Q,E_1}) are clearly infinite. However, since $f(r)\to f_{c}$ as $|\vec x|\to\infty$, it is reasonable to compute the charge and the energy of the Q-hole (Q-bulge) relative to the corresponding background solution (\ref{scalcondans}). Hence we define the renormalized charge and energy as follows,
\begin{eqnarray}\label{Q,E_2}
\begin{aligned}
Q_{ren}=Q-Q_{c}=2\omega\int f^{2}d^{d}x-2\omega\int f_{c}^{2}d^{d}x=2\omega\int \left(f^{2}-f_{c}^{2}\right)d^{d}x,\\
E_{ren}=E-E_{c}=\int\left(\omega^{2}\left(f^{2}-f_{c}^{2}\right)+\sum\limits_{i=1}^{d}\partial_{i}f\partial_{i}f+V(f)-V(f_{c})\right)d^{d}x.
\end{aligned}
\end{eqnarray}
Here $Q_{c}$ and $E_{c}$ are the scalar condensate charge and energy. The quantities (\ref{Q,E_2}) are finite as we will explicitly demonstrate below. Furthermore, they obey the following relation,
\begin{equation}\label{EQqholes}
E_{ren}=\omega Q_{ren}+\frac{2}{d}\int d^{d}x\,\partial_{i}f\partial_{i}f,
\end{equation}
which is analogous to that for Q-balls.\footnote{The proof of Eq.~(\ref{EQqholes}) is also fully analogous to that for Q-balls, the latter can be found in \cite{Gulamov:2013ema}.} Last, but not least, the following key property of Q-holes (Q-bulges) can be deduced from Eqs.~(\ref{Q,E_2}),
\begin{equation}\label{dedq}
\frac{dE_{ren}}{d\omega}=\omega\frac{dQ_{ren}}{d\omega}.
\end{equation}
The relation (\ref{dedq}) is also well known to be valid for Q-balls (with $Q_{ren}$ and $E_{ren}$ substituted by the genuine charge and energy of the Q-ball). This justifies the meaningfulness of our notions of $Q_{ren}$ and $E_{ren}$.

Note that since for Q-holes the inequality $f(\vec x)<f_{c}$ holds for all $|\vec{x}|<\infty$, the sign of $Q_{ren}$ is opposite to the sign of $\omega$ and the renormalized energy $E_{ren}$ is not positive definite. To prevent possible confusion, we stress again that $E_{ren}$ is defined with respect to the energy of the corresponding background solution and has no absolute meaning. Hence, unlike Q-balls, it is not possible to select a universal ground energy level from which one can count the energy of Q-holes (this reasoning holds for Q-bulges as well). Instead, the energy of each Q-hole (Q-bulge) must be renormalized in a unique way. As for Q-bulges, $\omega Q_{ren}>0$ in this case, leading to $E_{ren}>0$.

We would like to point out once again that, although $\frac{dU_{\omega}(f)}{df}\bigl|_{f=f_{c}}=0$, the original potential $V(f)$ may not have zero derivatives everywhere except the origin. Therefore, in general, the asymptotes of Q-holes (Q-bulges) do not approach any false vacuum state, contrary to what our intuition says about the properties of solitons.

\section{Explicit examples}\label{Explicit examples of Q-holes}
In this section we consider the model allowing for analytical investigation of the scalar condensate and Q-holes. For this purpose it is convenient to choose a simple piecewise-parabolic potential of the model \cite{Theodorakis:2000bz},
\begin{equation}\label{Potential}
V(\vert\phi\vert)=M^2\vert\phi\vert^2\theta\left(1-\dfrac{\vert\phi\vert^2}{v^2}\right)+M^2\left(\vert\phi\vert^2-2\epsilon v\vert\phi\vert+2\epsilon v^2\right)\theta\left(\dfrac{\vert\phi\vert^2}{v^2}-1\right),
\end{equation}
where $|\phi|=\sqrt{\phi^{*}\phi}$, $M^2>0$, $\epsilon>0$, $v>0$, $\theta$ is the Heaviside step function with the convention $\theta(0)=\frac{1}{2}$. The potential (\ref{Potential}) consists of two parabolic parts joined together at the point $\vert\phi\vert=v$. It possesses at least one minimum at $\vert\phi\vert=0$. It is easy to see that for $\epsilon<1$ there are no other minima, while for $\epsilon>1$ the second (local or global) minimum is located at $\vert\phi\vert=\epsilon v$. The potential (\ref{Potential}) can be generalized by using different masses for large and small values of $\vert\phi\vert$.

The potential (\ref{Potential}) does not admit the existence of Q-bulges.\footnote{One can supplement the scalar field potential (\ref{Potential}) by an additional parabolic part for the large values of $|\phi|$ to obtain Q-bulge solutions.} In principle, Q-holes and Q-bulges are of the same kind --- both solutions describe inhomogeneities in the scalar condensate and possess the same properties described by Eqs.~(\ref{Q,E_2})--(\ref{dedq}). Meanwhile, a possible negativity of $E_{ren}$ for Q-holes seems to be their peculiar feature, which makes their analysis more interesting. For this reason, we select Q-holes for the more detailed investigation.

\subsection{Scalar condensate and its stability}\label{subs31}
First, we aim to study the time-dependent scalar condensate in the $(d+1)$-dimensional space-time. The spatially homogeneous solutions of the equation of motion (\ref{3+1 EoM}) with the potential (\ref{Potential}) take the form
\begin{equation}
\phi=f_{c}\,e^{i\omega t}.
\end{equation}
For $0<f_{c}<v$, Eq.~(\ref{3+1 EoM}) gives $|\omega|=M$ and the magnitude $f_{c}$ of the condensate is independent of $\omega$. For $f_{c}>v$, Eq.~(\ref{3+1 EoM}) gives
\begin{equation}\label{fcpiecewise}
f_{c}=\frac{v\epsilon M^2}{M^2-\omega^2}.
\end{equation}
From this expression it follows that $0\leq|\omega|<M$. On the other hand, the condition $f_{c}>v$ implies that $\omega^{2}>M^{2}(1-\epsilon)$, if $\epsilon\leq 1$, and $\omega^{2}\ge 0$ otherwise. Combining these restrictions, we obtain the allowed region for $\omega$,
\begin{equation}\label{omegalimcond}
\textrm{max}[0;1-\epsilon]M^{2}\le\omega^{2}<M^{2}.
\end{equation}

Next we determine the charge and the energy of the condensate. When $0<f_{c}<v$, they take the form
\begin{eqnarray}
Q_{c}=\int\rho_{q}\,d^{d}x=\int 2Mf_{c}^{2}d^{d}x,\\ E_{c}=\int\rho_{e}\,d^{d}x=\int 2M^{2}f_{c}^{2}d^{d}x,
\end{eqnarray}
and for  $f_{c}>v$ we have
\begin{eqnarray}\label{condchargeeq22}
&&Q_{c}=\int \frac{2\omega v^{2}\epsilon^{2}M^4}{(M^2-\omega^2)^{2}}\,d^{d}x,\\
&&E_{c}=\int \frac{\epsilon v^{2}M^2}{(M^2-\omega^2)^{2}}\bigl(
(2-\epsilon)M^4+(3\epsilon-4)\omega^2M^2+2\omega^4\bigr)d^{d}x,
\end{eqnarray}
where $\omega$ is bounded by Eq.~(\ref{omegalimcond}). It is clear that the total charge and energy of the condensate are infinite due to the infinite volume of space.

We see that the theory contains two series of condensate solutions. The solutions of the series with $f_{c}<v$ allow to interpret them as collections of particles of mass $M$. Indeed, for these solutions $\rho_{e}=M\rho_{q}$. The solutions with $f_{c}>v$, despite being condensate, cannot be interpreted in this way.

Let us now examine the classical stability of the condensate under small fluctuations. It is clear that the solutions
\begin{equation}
\phi=f_{c}\,e^{iMt},\;\;\;0<f_{c}<v
\end{equation}
are classically stable --- the corresponding fluctuations satisfy the standard Klein-Gordon equation. In order to study the stability of the second series,
\begin{equation}\label{fluct}
\phi=\frac{v\epsilon M^2}{M^2-\omega^2}\,e^{i\omega t},\qquad \textrm{max}[0;1-\epsilon]\le\omega^{2}<M^{2},
\end{equation}
we write the scalar field in the form
\begin{equation}\label{pertback}
\phi=e^{i\omega t}\frac{v\epsilon M^2}{M^2-\omega^2}+e^{i\omega t}\left(a e^{ik_{0}t-i\vec k\vec x}+b e^{-ik_{0}t+i\vec k\vec x}\right),
\end{equation}
where $a$ and $b$ are complex constants and $\vec k=(k_{1},..,k_{d})$. Then we substitute this representation into the equation of motion for the scalar field and obtain a linearized equation for the fluctuations above the condensate solution. The stability (instability) of the condensate is manifested in the absence (existence) of the solutions of the linearized equation with imaginary $k_{0}$. Straightforward calculations give the following equation on $k_{0}$ and $\vec k$,
\begin{equation}\label{excitdispl}
\left(k_{0}^{2}-{\vec k}^{2}-M^{2}+\omega^{2}\right)\left(k_{0}^{2}-{\vec k}^{2}\right)-4\omega^{2}k_{0}^{2}=0,
\end{equation}
whose solutions are given by
\begin{equation}
k_{0}^{2}=\frac{M^{2}+3\omega^{2}+2{\vec k}^{2}\pm\sqrt{\left(M^{2}+3\omega^{2}\right)^{2}+16\omega^{2}{\vec k}^{2}}}{2}.
\end{equation}
Since
\begin{eqnarray}\nonumber
\left(M^{2}+3\omega^{2}+2{\vec k}^{2}\right)^{2}=\left(M^{2}+3\omega^{2}\right)^{2}+16\omega^{2}{\vec k}^{2}+4(M^{2}-\omega^{2}){\vec k}^{2}+4\left({\vec k}^{2}\right)^{2}\\\ge \left(M^{2}+3\omega^{2}\right)^{2}+16\omega^{2}{\vec k}^{2},
\end{eqnarray}
we obtain $k_{0}^{2}\ge 0$, i.e., the scalar condensate is stable under small fluctuations.\footnote{In the general case, the scalar condensate is stable (i.e., $k_{0}^{2}\ge 0$ for any $\vec k$) if $\frac{d^{2}V}{df^{2}}\bigl|_{f=f_{c}}-\frac{1}{f_{c}}
\frac{dV}{df}\bigl|_{f=f_{c}}\ge 0$, see \cite{Nugaev:2015rna}.}

\subsection{Q-holes in (1+1)-dimensional space-time}
Let us now study in detail Q-holes in the $(1+1)$-dimensional space-time. The corresponding solutions of Eq.~(\ref{3+1 EoM}) take the form
\begin{equation}\label{solution1d}
f(x)=\left\lbrack\begin{array}{l}
v\dfrac{\cosh\left(\sqrt{M^2-\omega^2}\,x\right)}{\cosh\left(\sqrt{M^2-\omega^2}\,X\right)},\qquad |x|<X,\vspace{0.2cm}\\
v\dfrac{\epsilon M^2}{M^2-\omega^2}-v\left(\dfrac{\epsilon M^2}{M^2-\omega^2}-1\right)e^{\sqrt{M^2-\omega^2}(X-|x|)},\qquad |x|\ge X,
\end{array}\right.
\end{equation}
where
\begin{equation}\label{X1d}
X=\frac{1}{\sqrt{M^2-\omega^2}}\,\textrm{arctanh}\left(\frac{(\epsilon-1)M^2+\omega^2}{M^2-\omega^2}\right)
\end{equation}
defines the matching point at which $f(X)=v$. Since the argument of the inverse hyperbolic tangent should be less than unity and not less than zero, from Eq.~(\ref{X1d}) one finds
\begin{equation}\label{restromegaqholes}
\left(1-\epsilon\right)M^{2}\le\omega^{2}<\left(1-\frac{\epsilon}{2}\right)M^{2}.
\end{equation}
Note that the r.h.s. of the inequality (\ref{restromegaqholes}) can be obtained using the mechanical analogy.
It is clear from Fig.~\ref{plotUqholes} that if $U_{\omega}(f_{c})\ge 0$, then the particle will never reach the top of the effective potential. Hence the condition $\omega^{2}f_{c}^{2}-V(f_{c})<0$ holds, and using Eqs.~(\ref{fcpiecewise}) and (\ref{Potential}), we deduce $\omega^{2}<\left(1-\frac{\epsilon}{2}\right)M^{2}$.

From Eq.~(\ref{restromegaqholes}) it follows that Q-holes do not exist if $\epsilon\ge2$, while for $1<\epsilon<2$ the l.h.s. of Eq.~(\ref{restromegaqholes}) can be replaced by $0\leq \omega^2$. Hence the allowed region for $\omega$ in the theory with the potential (\ref{Potential}) takes the form (cf. Eq.~(\ref{omegalimcond}))
\begin{equation}
\textrm{max}[0;1-\epsilon]M^{2}\le\omega^{2}<\left(1-\frac{\epsilon}{2}\right)M^{2}.
\end{equation}

\begin{figure}[ht]
\center{\includegraphics[width=0.7\linewidth]{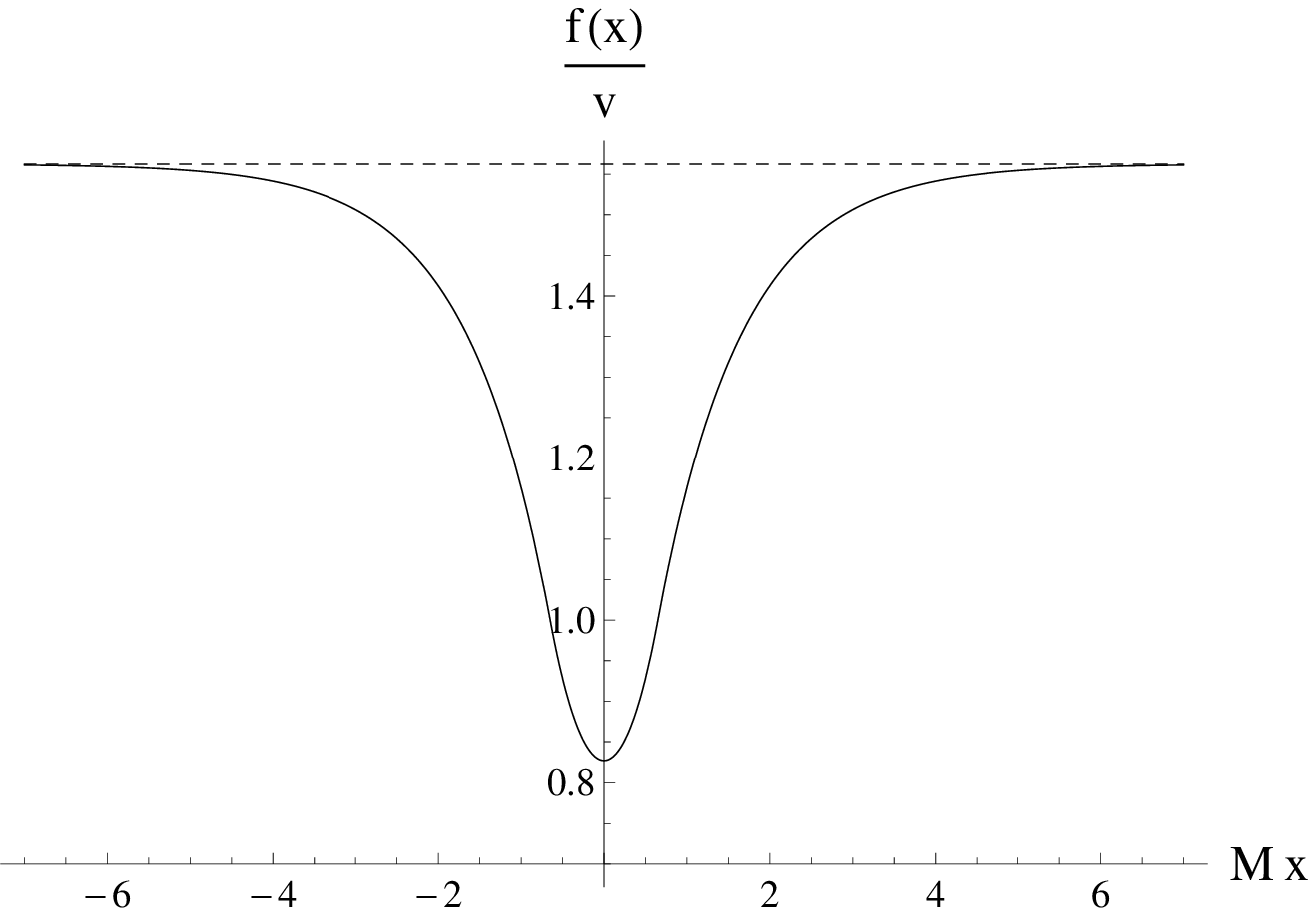}}
\caption{Q-hole solution in the $(1+1)$-dimensional space-time. Here $\epsilon=1.5$, $\omega=0.2 M$, the dashed line stands for the scalar condensate $f_{c}$.}
\label{qhole1d}
\end{figure}

An example of the Q-hole solution is presented in Fig.~\ref{qhole1d}. We observe that $f(x)\to f_{c}>v$ as $x\to\pm\infty$. It is not difficult to show that Q-holes with the asymptotes $\lim\limits_{x\to\pm\infty}f(x)\to f_{c}<v$ do not exist in the theory with the potential (\ref{Potential}).\footnote{This statement remains true in the $(d+1)$-dimensional case.}

\begin{figure}[!ht]
\begin{minipage}{0.49\linewidth}
\center{\includegraphics[width=0.9\linewidth]{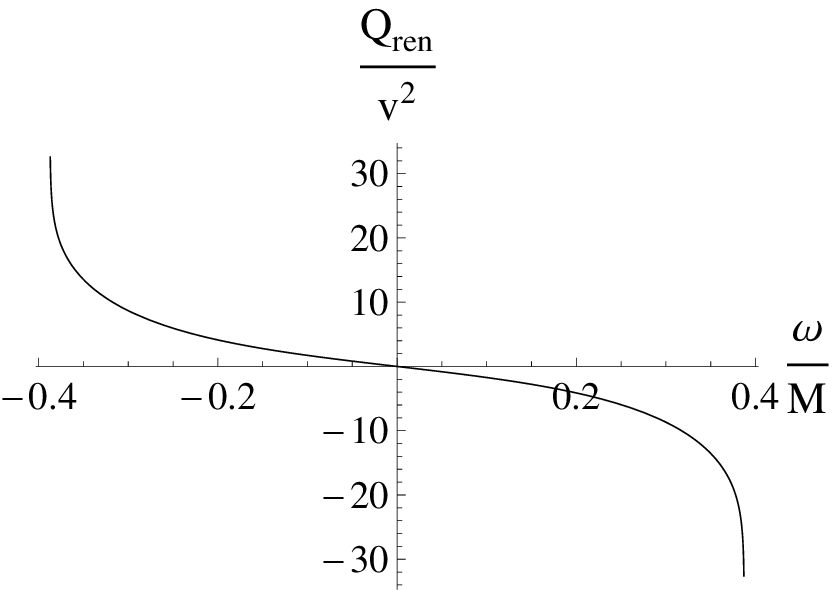}}
\end{minipage}
\begin{minipage}{0.49\linewidth}
\center{\includegraphics[width=0.9\linewidth]{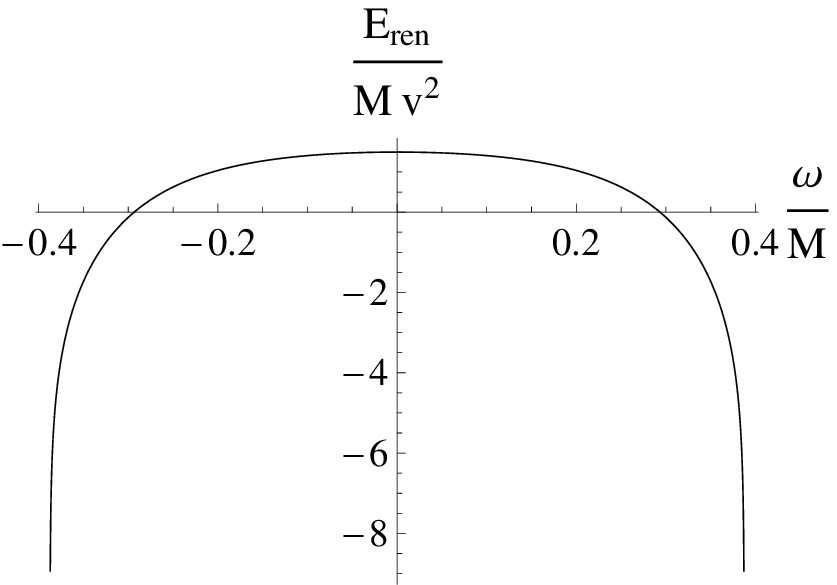}}
\end{minipage}
\caption{$Q_{ren}(\omega)$ and $E_{ren}(\omega)$ for $\epsilon=1.7$ in the $(1+1)$-dimensional case.}
\label{QE1d-1}
\end{figure}

\begin{figure}[!ht]
\begin{minipage}{0.49\linewidth}
\center{\includegraphics[width=0.9\linewidth]{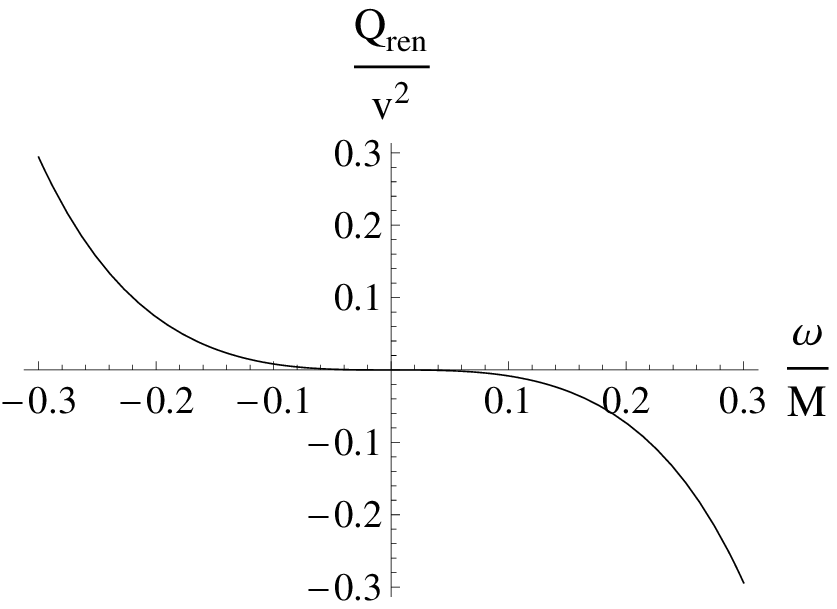}}
\end{minipage}
\begin{minipage}{0.49\linewidth}
\center{\includegraphics[width=0.9\linewidth]{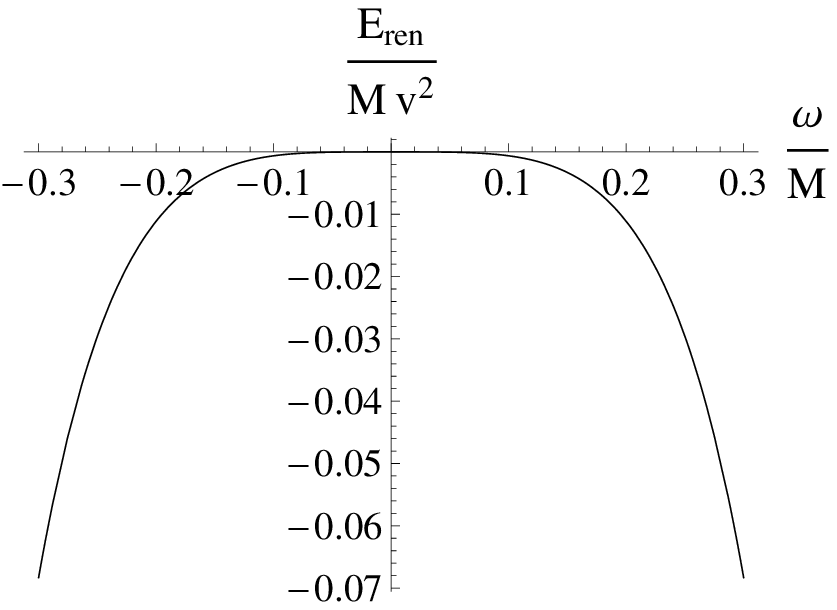}}
\end{minipage}
\caption{$Q_{ren}(\omega)$ and $E_{ren}(\omega)$ for $\epsilon=1$ in the $(1+1)$-dimensional case.}
\label{QE1d-2}
\end{figure}

\begin{figure}[!ht]
\begin{minipage}{0.49\linewidth}
\center{\includegraphics[width=0.9\linewidth]{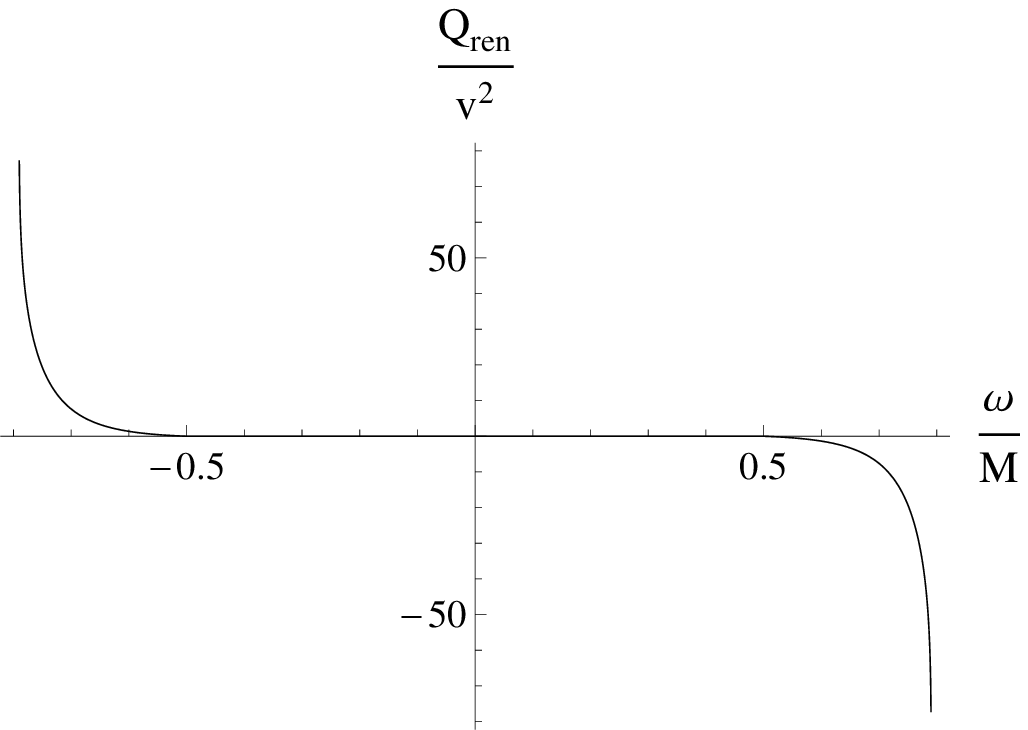}}
\end{minipage}
\begin{minipage}{0.49\linewidth}
\center{\includegraphics[width=0.9\linewidth]{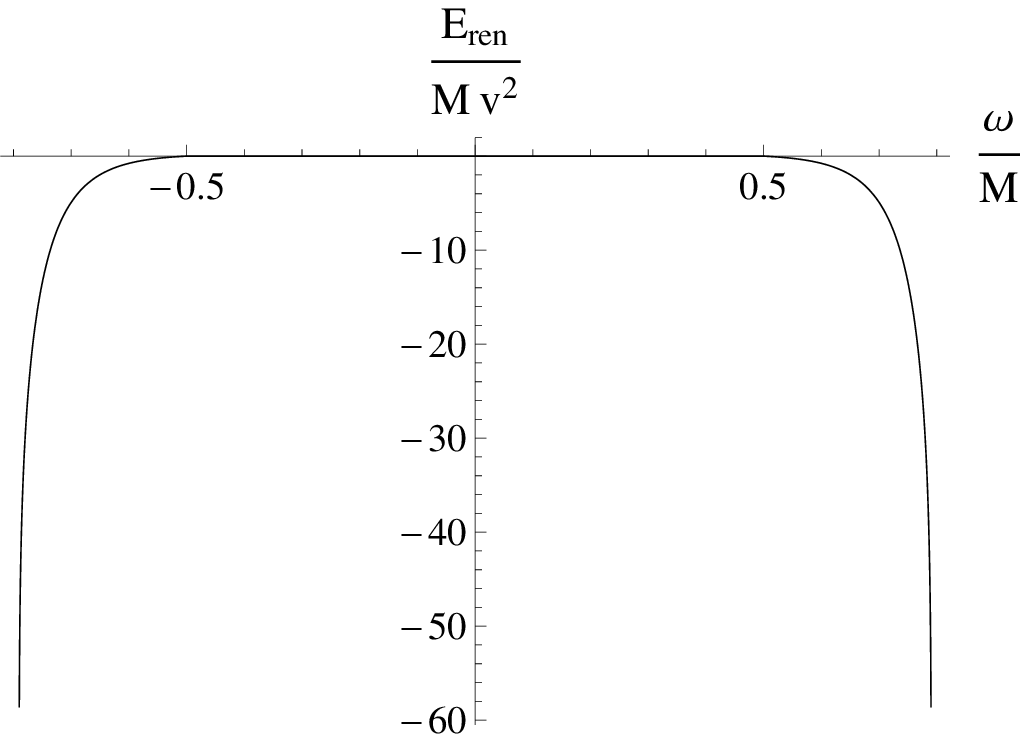}}
\end{minipage}
\caption{$Q_{ren}(\omega)$ and $E_{ren}(\omega)$ for $\epsilon=0.75$ in the $(1+1)$-dimensional case.}
\label{QE1d-3}
\end{figure}

For the renormalized charge and energy we get
\begin{eqnarray}\nonumber
Q_{ren}=\frac{2\omega v^{2}}{\left(M^{2}-\omega^{2}\right)^{\frac{5}{2}}}\Biggl(\Bigl((2\epsilon-3\epsilon^{2})M^{4}-2\epsilon M^{2}\omega^{2}\Bigr)\textrm{arctanh}\left(\frac{(\epsilon-1)M^2+\omega^2}{M^2-\omega^2}\right)\\-3\epsilon M^{2}\left((\epsilon-1)M^{2}+\omega^{2}\right)\Biggr),\label{Q1d}\\ \nonumber
E_{ren}=\frac{2v^{2}}{\left(M^{2}-\omega^{2}\right)^{\frac{5}{2}}}\Biggl(\Bigl(\epsilon^{2}M^{4}(M^{2}-4\omega^{2})+\epsilon M^{2}(M^{2}-\omega^{2})(4\omega^{2}-2M^{2})\Bigr)\\\times\textrm{arctanh}\left(\frac{(\epsilon-1)M^2+\omega^2}{M^2-\omega^2}\right)+
\epsilon M^{2}(M^{2}-4\omega^{2})\left((\epsilon-1)M^{2}+\omega^{2}\right)\Biggr)\label{E1d}.
\end{eqnarray}
Let us mention some properties of $Q_{ren}$ and $E_{ren}$ following from Eqs.~(\ref{Q1d}) and (\ref{E1d}).
\begin{enumerate}
\item $|Q_{ren}|\to\-\infty$ and $|E_{ren}|\to\-\infty$ as $|\omega|\to M\sqrt{1-{\epsilon}/{2}}$. Indeed, in this case $X\to\infty$, whereas $f(x)$ tends to the vacuum solution $f\equiv 0$ at $|x|<X$.
\item For $\omega=0$ (if allowed, i.e., if $\epsilon>1$), $Q_{ren}=0$ and $E_{ren}>0$ due to Eq.~(\ref{EQqholes}).
\item For $\omega=M\sqrt{1-{\epsilon}}$ and $\epsilon<1$, we get $Q_{ren}=0$ and $E_{ren}=0$. Indeed, in this case $X=0$ and $f(x)\equiv f_{c}=v$.
\end{enumerate}
Some typical examples of $Q_{ren}(\omega)$ and $E_{ren}(\omega)$ dependencies are presented in Figs.~\ref{QE1d-1}--\ref{QE1d-3}. We see that the renormalized energy $E_{ren}$ can take positive as well as negative and zero values. As was explained in Section~\ref{Time-dependent scalar condensate and Q-holes}, this result is expected and should not surprise. As a useful check of validity of our calculations, one can show numerically that Eq.~(\ref{dedq}) fulfills for $Q_{ren}$ and $E_{ren}$ given by Eqs.~(\ref{Q1d}) and (\ref{E1d}).

Let us pause here to make a general comment on a choice of regularization scheme for Q-holes (and Q-bulges). Eqs.~(\ref{Q,E_2}) give a natural way to obtain finite values for the charge and energy of the Q-hole (Q-bulge). The corresponding quantities $Q_{ren}$ and $E_{ren}$ satisfy all the relations they are expected to satisfy as the ``charge'' and the ``energy'' of the soliton. We can expect, therefore, that any consistent regularization must lead to the same expressions for $Q_{ren}$ and $E_{ren}$. One such scheme corresponds to putting the system in a box of size $2L$ (with the natural boundary conditions $f(-L)=f(L)$ and $\frac{df}{dx}\bigr|_{x=-L}=\frac{df}{dx}\bigr|_{x=L}=0$), computing $Q_{ren}$ and $E_{ren}$ as differences of finite quantities, and taking the limit $L\rightarrow\infty$. This procedure endows Eqs.~(\ref{Q,E_2}) with the precise meaning. We conclude that the negativity of $E_{ren}$ at some $\omega$ is an inherent property of Q-holes and not a consequence of a particular choice of regularization.

\subsection{Q-holes in (3+1)-dimensional space-time}
The analysis of Q-holes in three spatial dimensions lies closely to that in the $(1+1)$-dimensional case. The spherically symmetric ansatz for the scalar field reads as follows,
\begin{equation}
\phi(t,\vec x)=f(r)e^{i\omega t},
\end{equation}
where $r=\sqrt{\vec x^{2}}$. The solutions to the equation of motion (\ref{3+1 EoM}) with the potential (\ref{Potential}) take the form
\begin{equation}\label{solution3d}
f(r)=\left\lbrack\begin{array}{l}
v\dfrac{\sinh\left(\sqrt{M^2-\omega^2}\,r\right)}{\sinh\left(\sqrt{M^2-\omega^2}\,R\right)}\dfrac{R}{r},\qquad r<R, \vspace{0.2cm}\\
v\dfrac{\epsilon M^{2}}{M^{2}-\omega^2}-v\left(\dfrac{\epsilon M^{2}}{M^{2}-\omega^2}-1\right)\dfrac{R}{r}\,e^{\sqrt{M^2-\omega^2}\,(R-r)},\qquad r\ge R,
\end{array}\right.
\end{equation}
where $R$ is defined by
\begin{equation}\label{Rqholes3d}
\sqrt{M^2-\omega^2}\coth\left(\sqrt{M^2-\omega^2}\,R\right)=\frac{1}{R}+\left(\frac{\epsilon M^{2}}{M^{2}-\omega^2}-1\right)\left(\sqrt{M^2-\omega^2}+\frac{1}{R}\right).
\end{equation}
Contrary to the $(1+1)$-dimensional case, the latter equation has no analytical solutions for $R$. However, it can be solved numerically.

\begin{figure}[ht]
\center{\includegraphics[width=0.7\linewidth]{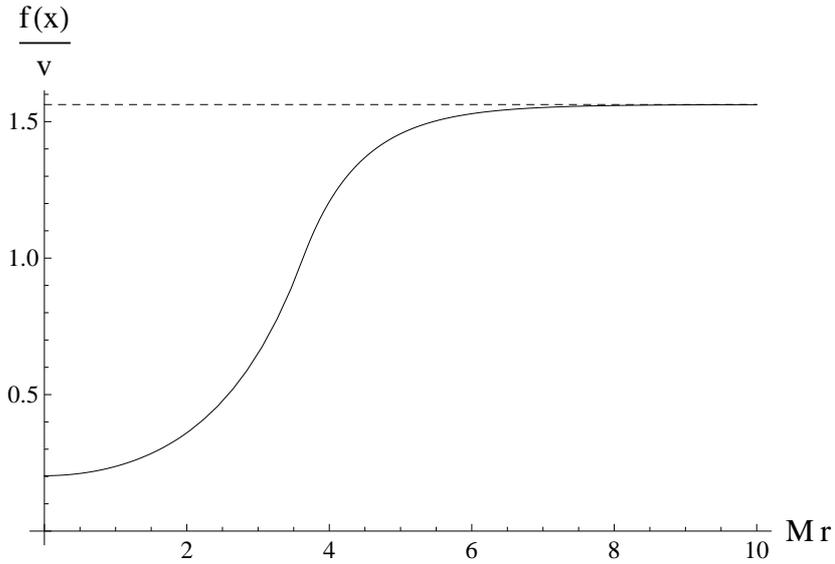}}
\caption{Q-hole solution in the $(3+1)$-dimensional space-time. Here $\epsilon=1.5$, $\omega=0.2 M$, the dashed line stands for the scalar condensate $f_{c}$.}
\label{qhole3d}
\end{figure}

Acting exactly as in the $(1+1)$-dimensional case, with the use of the mechanical analogy one can obtain the relation $\omega^{2}<\left(1-\frac{\epsilon}{2}\right)M^{2}$ leading together with Eq.~(\ref{omegalimcond}) to
\begin{equation}\label{restr3d}
\textrm{max}[0;1-\epsilon]M^{2}\le\omega^{2}<\left(1-\frac{\epsilon}{2}\right)M^{2}.
\end{equation}
Again, from the latter equation it follows that Q-holes do not exist if $\epsilon\ge 2$. Although Eq.~(\ref{Rqholes3d}) is more complicated than Eq.~(\ref{X1d}) in the $(1+1)$-dimensional case, the restriction (\ref{restr3d}) can also be obtained directly from Eq.~(\ref{Rqholes3d}), see Appendix~A for details. It is interesting to note that both in $1+1$ and $3+1$ dimensions the model admits Q-ball solutions.\footnote{The explicit $(3+1)$-dimensional solutions can be found in \cite{Theodorakis:2000bz}.} As can be easily shown using the mechanical analogy, they exist if
\begin{equation}
\left(1-\frac{\epsilon}{2}\right)M^{2}<\omega^{2}<M^{2}.
\end{equation}
An example of the Q-hole solution in $3+1$ dimensions is presented in Fig.~\ref{qhole3d}.

The renormalized charge and energy are given by
\begin{eqnarray}\nonumber
Q_{ren}=\frac{v^{2}}{M^{2}}8\pi\tilde\omega\left(\frac{\tilde R}{2(1-{\tilde\omega}^{2})}
\left({\tilde R}^{2}(1-{\tilde\omega}^{2})+\frac{1}{4}-\left(\tilde f_{c}+\sqrt{1-{\tilde\omega}^{2}}\tilde R ({\tilde f_{c}}-1)-\frac{1}{2}\right)^{2}\right)\right.\\\left.
-\frac{{\tilde R}^{3}{\tilde f_{c}}^{2}}{3}+\frac{{\tilde R}^{2}({\tilde f_{c}}-1)^{2}}{2\sqrt{1-{\tilde\omega}^{2}}}-2{\tilde f_{c}}({\tilde f_{c}}-1)
\left(\frac{{\tilde R}^{2}}{\sqrt{1-{\tilde\omega}^{2}}}+\frac{{\tilde R}}{1-{\tilde\omega}^{2}}\right)\right),\label{Q3d}
\\\nonumber
E_{ren}=\frac{v^{2}}{M}8\pi\left(\frac{(4{\tilde\omega}^{2}-1)\tilde R}{6(1-{\tilde\omega}^{2})}
\left({\tilde R}^{2}(1-{\tilde\omega}^{2})+\frac{1}{4}-\left(\tilde f_{c}+\sqrt{1-{\tilde\omega}^{2}}\tilde R ({\tilde f_{c}}-1)-\frac{1}{2}\right)^{2}\right)\right.\\\left.
-\frac{{\tilde\omega}^{2}{\tilde R}^{3}{\tilde f_{c}}^{2}}{3}+\frac{(4{\tilde\omega}^{2}-1){\tilde R}^{2}({\tilde f_{c}}-1)^{2}}{6\sqrt{1-{\tilde\omega}^{2}}}-\frac{(7{\tilde\omega}^{2}-1){\tilde f_{c}}({\tilde f_{c}}-1)}{3}
\left(\frac{{\tilde R}^{2}}{\sqrt{1-{\tilde\omega}^{2}}}+\frac{{\tilde R}}{1-{\tilde\omega}^{2}}\right)\right),\label{E3d}
\end{eqnarray}
where the notations $\tilde\omega=\frac{\omega}{M}$, $\tilde R=MR$ and $\tilde f_{c}=\frac{\epsilon M^2}{M^2-\omega^2}$ are introduced to shorten the formulas.

Some general properties of $Q_{ren}$ and $E_{ren}$ defined by Eqs.~(\ref{Q3d}) and (\ref{E3d}) are in order.
\begin{enumerate}
\item $|Q_{ren}|\to\-\infty$ and $|E_{ren}|\to\-\infty$ as $|\omega|\to M\sqrt{1-{\epsilon}/{2}}$. Indeed, in this case $R\to\infty$, whereas $f(r)$ tends to the vacuum solution $f\equiv 0$ at $r<R$.
\item For $\omega=0$ (if allowed, i.e., if $\epsilon>1$), $Q_{ren}=0$ and $E_{ren}>0$ due to Eq.~(\ref{EQqholes}). This is an expected result for the solution with $\omega=0$, which is just a sphaleron.
\item For $\omega=M\sqrt{1-{\epsilon}}$ and $\epsilon<1$, we have $Q_{ren}=0$ and $E_{ren}=0$. Indeed, in this case $R=0$ and $f(r)\equiv f_{c}=v$.
\end{enumerate}

\begin{figure}[!ht]
\begin{minipage}{0.49\linewidth}
\center{\includegraphics[width=0.9\linewidth]{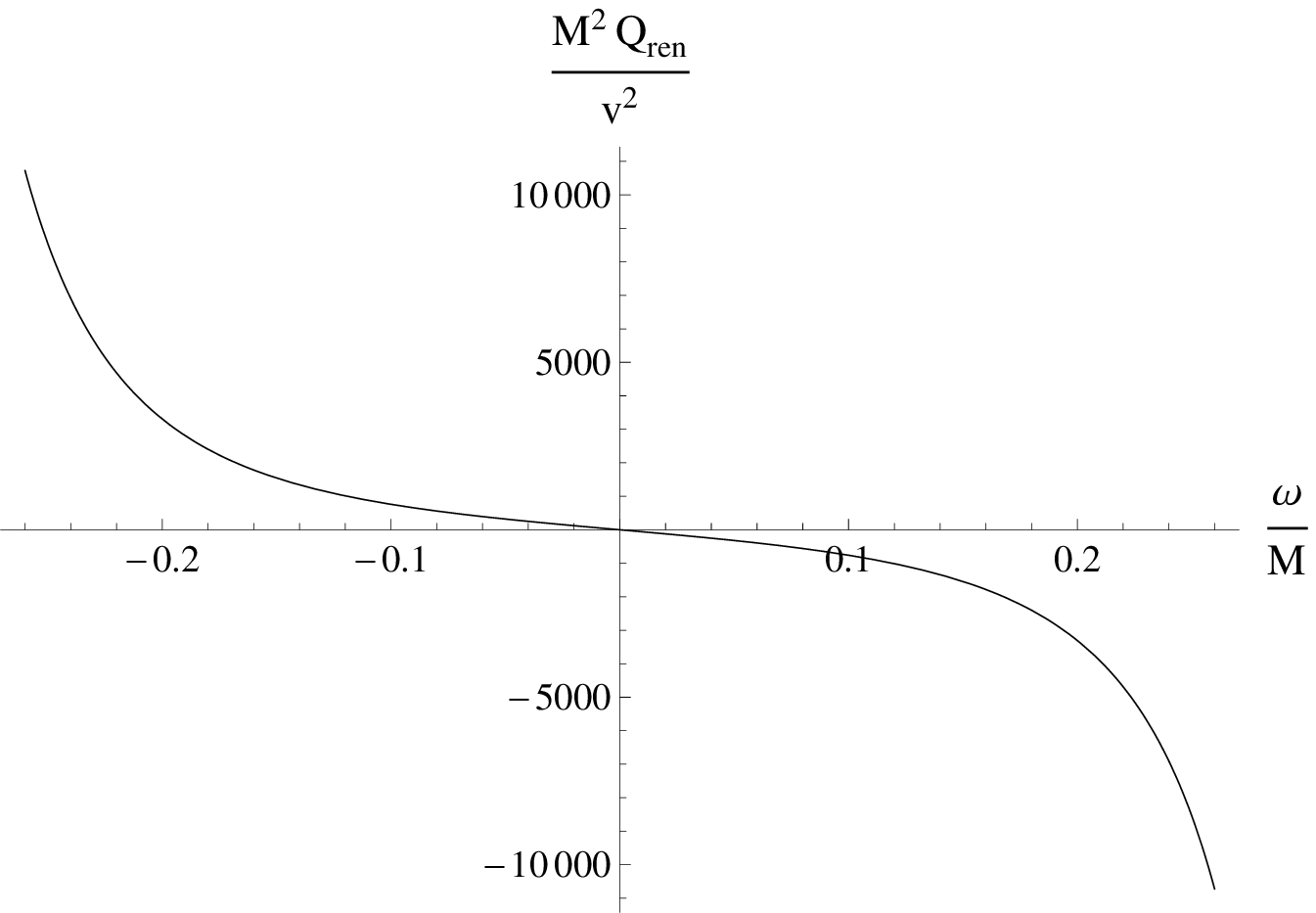}}
\end{minipage}
\begin{minipage}{0.49\linewidth}
\center{\includegraphics[width=0.9\linewidth]{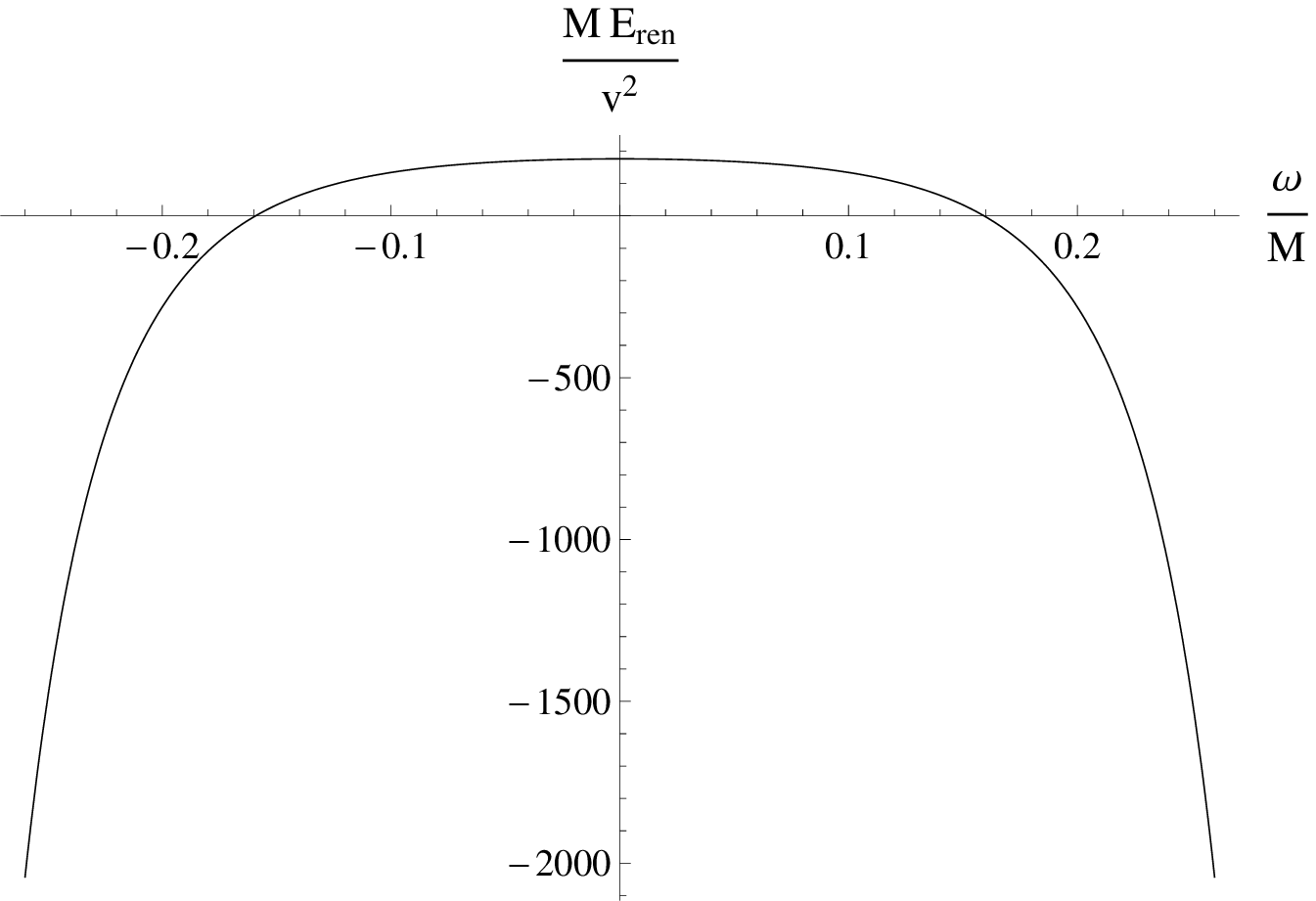}}
\end{minipage}
\caption{$Q_{ren}(\omega)$ and $E_{ren}(\omega)$ for $\epsilon=1.7$ in the $(3+1)$-dimensional case.}
\label{QE3d-1}
\end{figure}
\begin{figure}[!ht]
\begin{minipage}{0.49\linewidth}
\center{\includegraphics[width=0.9\linewidth]{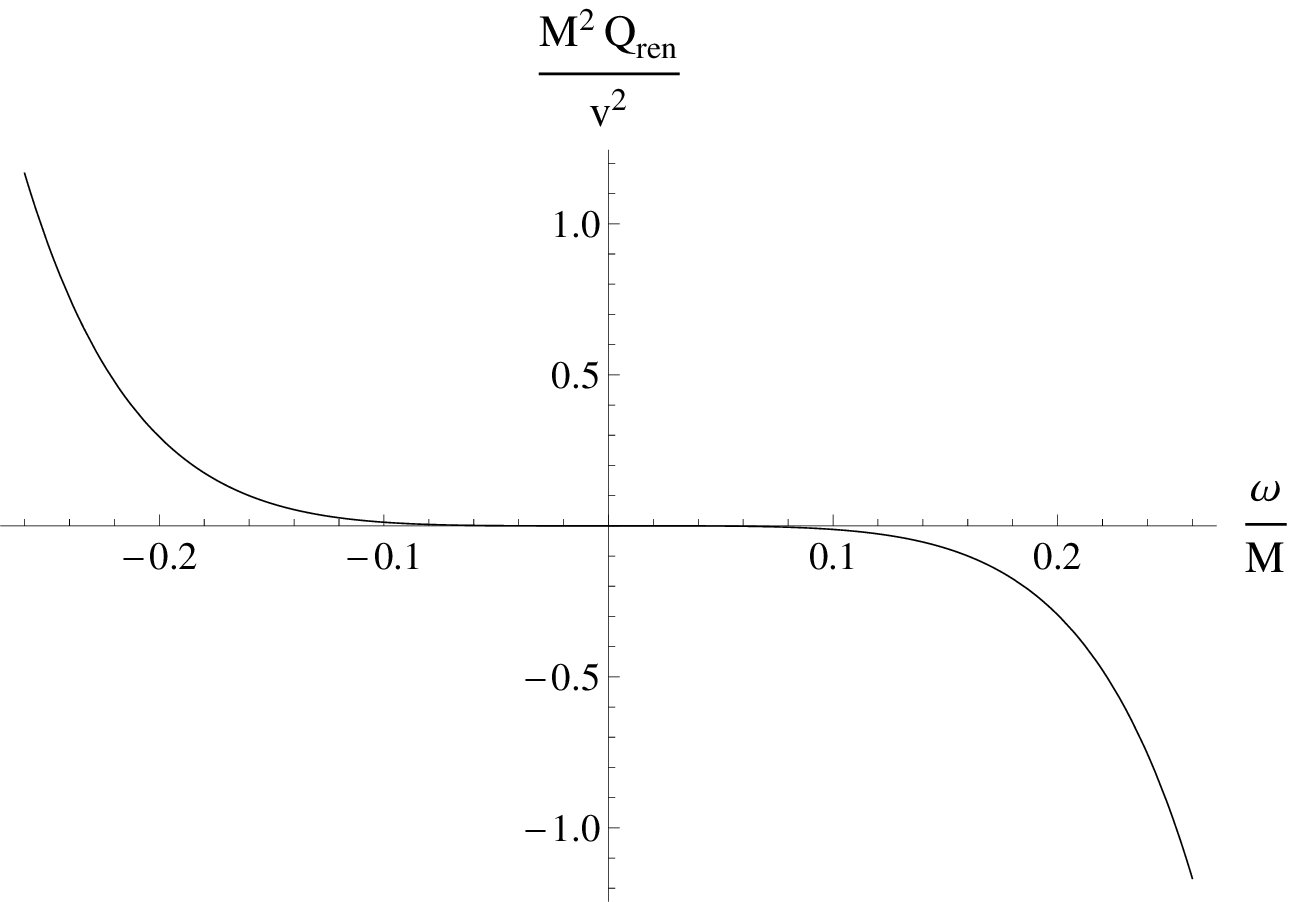}}
\end{minipage}
\begin{minipage}{0.49\linewidth}
\center{\includegraphics[width=0.9\linewidth]{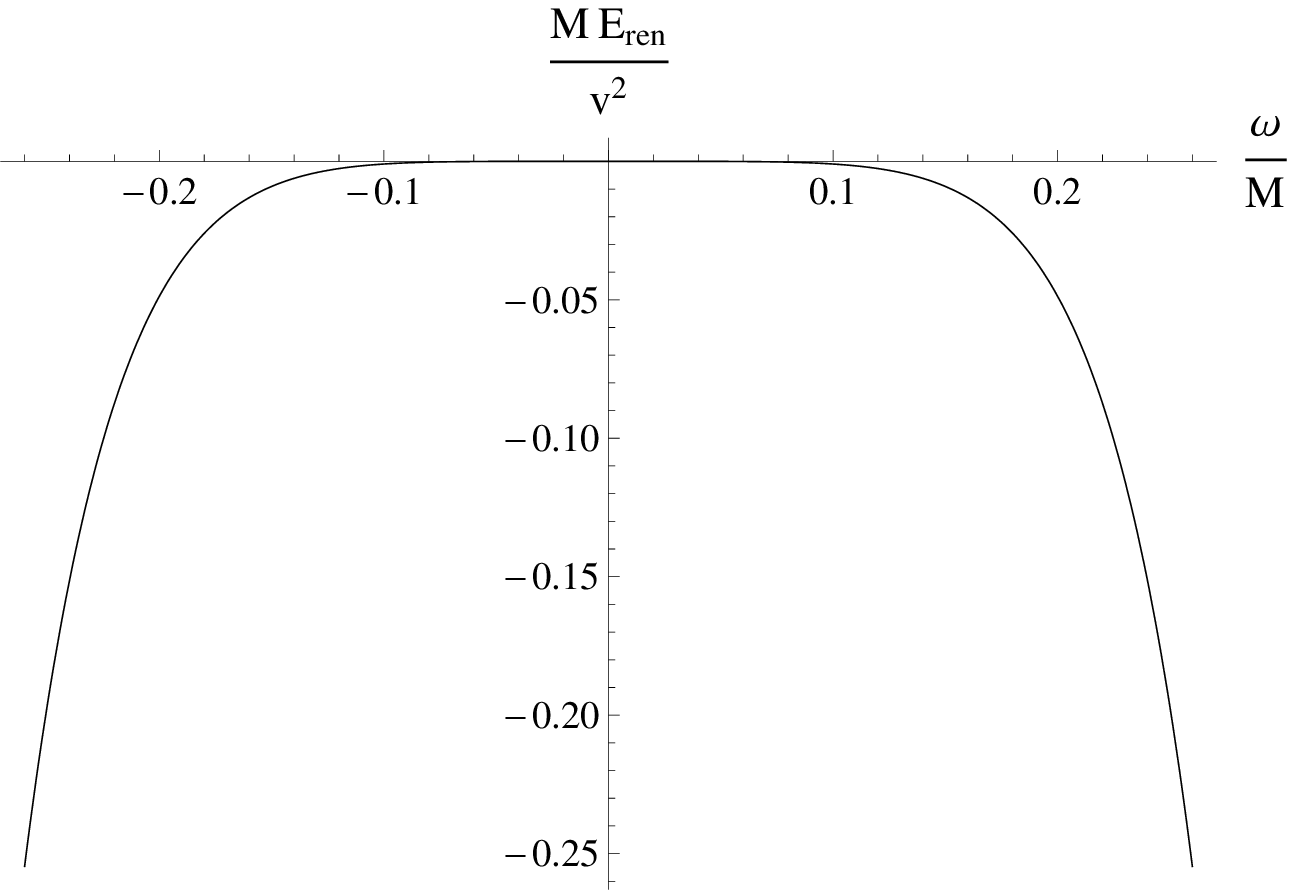}}
\end{minipage}
\caption{$Q_{ren}(\omega)$ and $E_{ren}(\omega)$ for $\epsilon=1$ in the $(3+1)$-dimensional case.}
\label{QE3d-2}
\end{figure}
\begin{figure}[!ht]
\begin{minipage}{0.49\linewidth}
\center{\includegraphics[width=0.9\linewidth]{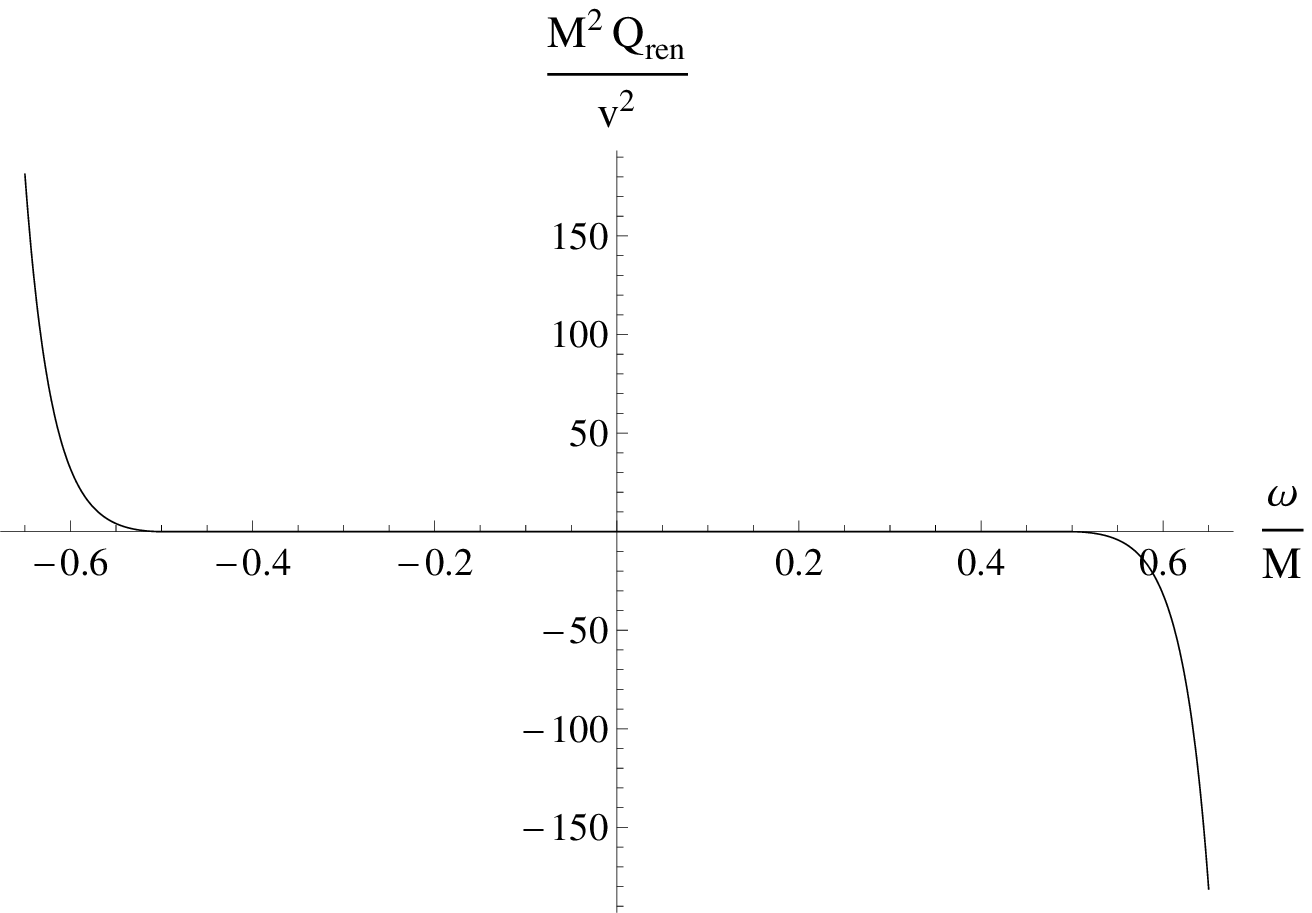}}
\end{minipage}
\begin{minipage}{0.49\linewidth}
\center{\includegraphics[width=0.9\linewidth]{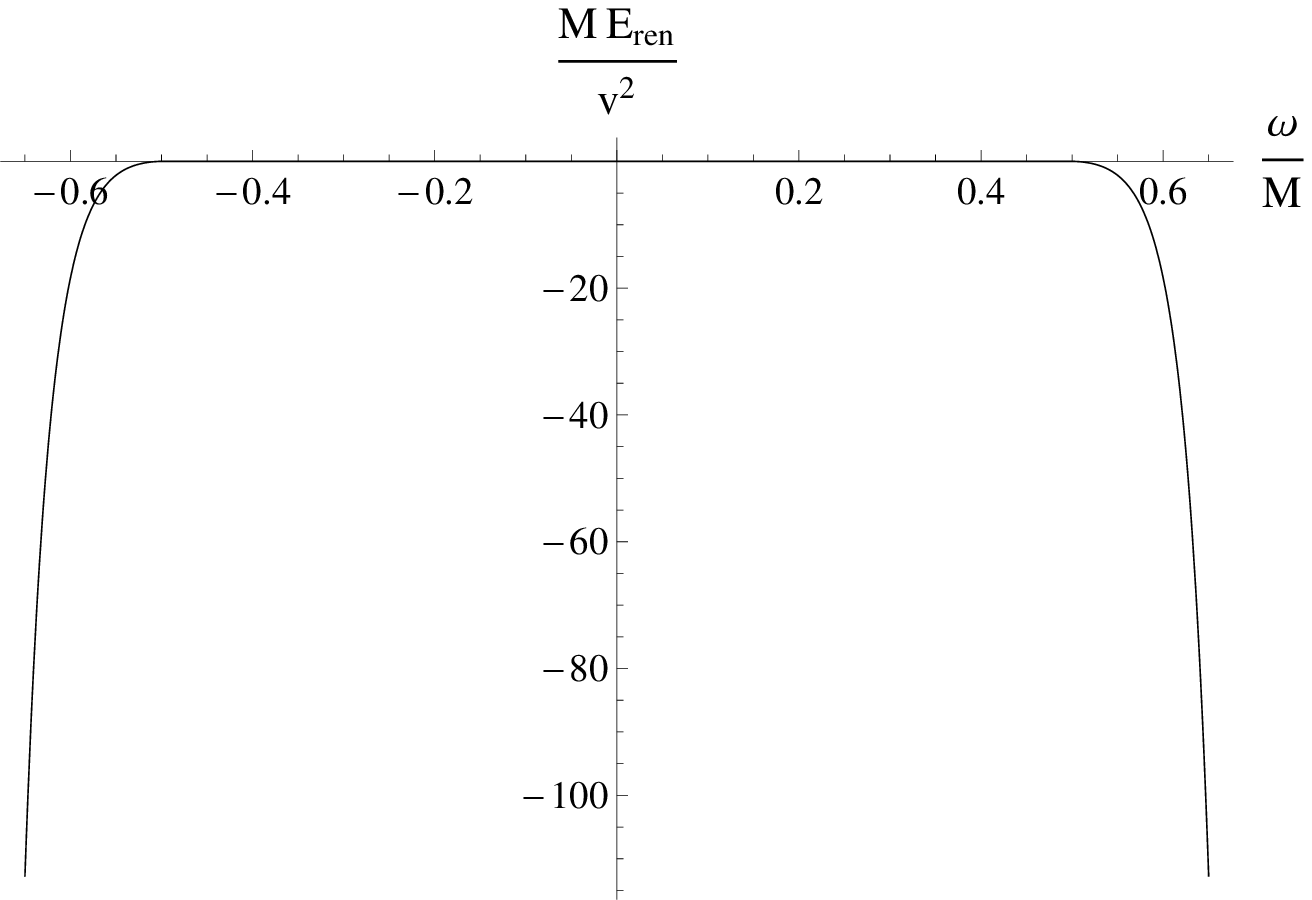}}
\end{minipage}
\caption{$Q_{ren}(\omega)$ and $E_{ren}(\omega)$ for $\epsilon=0.75$ in the $(3+1)$-dimensional case.}
\label{QE3d-3}
\end{figure}
Some typical examples of the $Q_{ren}(\omega)$ and $E_{ren}(\omega)$ dependencies are presented in Figs.~\ref{QE3d-1}--\ref{QE3d-3}. Again, we see that the renormalized energy $E_{ren}$ can be positive, negative\footnote{The negativity of $E_{ren}$ poses the question about possible spontaneous creation of Q-holes. However, at the moment it is not clear what can carry the rest of the charge and energy (in comparison with the condensate, i.e., $-Q_{ren}$ and $-E_{ren}$), taking into account the nonstandard form of excitations above the condensate, see Subsection~\ref{subs31}. This problem will be briefly discussed below.} or zero. As in the $(1+1)$-dimensional case, one can check numerically that the relation (\ref{dedq}) fulfills for $Q_{ren}$ and $E_{ren}$ given by Eqs.~(\ref{Q3d}) and (\ref{E3d}).

\section{Classical instability of Q-holes and Q-bulges}
There is a well-known classical stability criterion for Q-balls \cite{Friedberg:1976me,LeePang}, which states that Q-balls with $\frac{dQ}{d\omega}<0$ are classically stable.\footnote{Speaking more precisely, this condition must be supplemented by an additional requirement on the number of negative eigenvalues of some operator, which arises when considering perturbations above the Q-ball.} It is easy to see that the method used to obtain this criterion (as well as the similar approach used in \cite{Q-criterion} for obtaining the stability criterion for NSE) cannot be generalized straightforwardly to the case of Q-holes and Q-bulges with the (renormalized) charge and energy given by $Q_{ren}$ and $E_{ren}$. Indeed, contrary to the case of ordinary Q-balls, whose asymptotics at $r\rightarrow\infty$ are the same for any value of $\omega$, for Q-holes and Q-bulges the asymptotic behavior is different for different $\omega$. Moreover, their total charge and energy are infinite. Despite these obstacles, one can give some arguments in favor of classical instability of these solitons, to which we now proceed.

As was mentioned in Section~3.2, one can put the system in a box of finite size and regard $Q_{ren}$ as a difference $Q-Q_c$ between the charges of soliton and condensate computed in this box. The box implies boundary conditions to be imposed on the solutions. For example, in $1+1$ dimensions with the size of the spatial dimension $2L$, one can demand the periodic boundary conditions $f(-L)=f(L)$ and $\frac{df}{dx}\bigr|_{x=-L}=\frac{df}{dx}\bigr|_{x=L}=0$. The solution obeying these conditions can be easily obtained for the potential (\ref{Potential}). It is also easy to check that this solution does not have nodes and tends to (\ref{solution1d}) for $L\to\infty$. Hence one can expect that, as long as the characteristic scale of the soliton $l$ is much smaller than the size of the box $L$, the value of $Q_{ren}$ lies very close to its limit at $L\rightarrow\infty$. Since $\frac{Q_{ren}}{Q}\ll 1$ for $l\ll L$, for such solutions one can write $Q\approx Q_{c}$.

Since $Q$ is finite in the box, we have no obstacles in the derivation of the Q-criterion \cite{Friedberg:1976me,LeePang} that would forbid us to apply it to our case. Choosing the size of the box to be sufficiently large, we have
\begin{equation}\label{dQdQc}
\frac{dQ}{d\omega}\approx\frac{dQ_{c}}{d\omega}.
\end{equation}
We see that the sign of $\frac{dQ}{d\omega}$, determining the (in)stability of the solution, follows from the sign of $\frac{dQ_{c}}{d\omega}$. We now ask what the sign of $\frac{dQ_{c}}{d\omega}$ is.

We will be interested in the case of a stable scalar condensate, like the one in Eq.~(\ref{fluct}), for which the condition
\begin{equation}\label{condstabcond}
\frac{d^{2}V}{df^{2}}\biggl|_{f=f_{c}}-\frac{1}{f_{c}}
\frac{dV}{df}\biggl|_{f=f_{c}}\ge 0
\end{equation}
holds \cite{Nugaev:2015rna}. The condensate charge is $Q_{c}=2\omega f_{c}^{2}V^{(d)}$, where $V^{(d)}\sim L^{d}$ is the space volume. Thus,
\begin{equation}\label{dqcdomega}
\frac{dQ_{c}}{d\omega}=2V^{(d)}\left(f_{c}^{2}+2\omega f_{c}\frac{df_{c}}{d\omega}\right).
\end{equation}
Now, differentiating Eq.~(\ref{fcdef}) with respect to $\omega$, using Eq.~(\ref{Upotdef}) and multiplying the result by $\frac{df_{c}}{d\omega}$, we get
\begin{equation}\label{auxilfc}
2\omega f_{c}\frac{df_{c}}{d\omega}=\frac{1}{2}\left(\frac{d^{2}V}{df^{2}}\biggl|_{f=f_{c}}-\frac{1}{f_{c}}
\frac{dV}{df}\biggl|_{f=f_{c}}\right)\left(\frac{df_{c}}{d\omega}\right)^{2}.
\end{equation}
Substituting Eq.~(\ref{auxilfc}) into Eq.~(\ref{dqcdomega}), we arrive at
\begin{equation}\label{dQdomegacondensate}
\frac{dQ_{c}}{d\omega}=2V^{(d)}\left(f_{c}^{2}+\frac{1}{2}\left(\frac{d^{2}V}{df^{2}}\biggl|_{f=f_{c}}-\frac{1}{f_{c}}
\frac{dV}{df}\biggl|_{f=f_{c}}\right)\left(\frac{df_{c}}{d\omega}\right)^{2}\right).
\end{equation}
We see that whenever the condensate stability criterion (\ref{condstabcond}) fulfills, the relation $\frac{dQ_{c}}{d\omega}>0$ holds (for Eq.~(\ref{condchargeeq22}), this can be checked explicitly). The latter inequality means that solutions for which Eq.~(\ref{dQdQc}) holds may be (and probably are) classically {\em unstable}.\footnote{Since $\frac{1}{V^{(d)}}\frac{dQ_{c}}{d\omega}\ge 2f_{c}^{2}>0$ for the classically stable condensate (see Eq.~(\ref{dQdomegacondensate})), for any finite $\frac{dQ_{ren}}{d\omega}$ we can always take such a large spatial volume $V^{(d)}$ that $\frac{dQ}{d\omega}>0$.} Namely, most probably there exists a mode of the form $\varphi(\vec x)e^{i\omega t}e^{\gamma t}$ among the excitations above the Q-hole or Q-bulge, where $\gamma$ is a real constant. Finally, it is reasonable to suppose that Q-holes and Q-bulges remain classically unstable in the limit $V^{(d)}\to\infty$.

An important remark is in order here. The inequality $\frac{dQ_{c}}{d\omega}>0$ holds for the classically {\em stable} condensate, which may seem confusing as the stability criterion for the Q-solitons dictates $\frac{dQ}{d\omega}<0$. However, there is no contradiction here --- the method of \cite{Friedberg:1976me,LeePang} is based on the existence of a negative eigenvalue of the already mentioned operator (more precisely, it is the operator $h_{R}$ defined by Eq.~(3.39) of \cite{LeePang}), which arises when one considers perturbations above the Q-ball. One can check that for the scalar condensate satisfying Eq.~(\ref{condstabcond}), this operator can not have negative eigenvalues at all. The latter makes this method and, consequently, the Q-criterion  not applicable to the spatially-homogeneous solutions.

The conclusion about classical instability of Q-holes and Q-bulges is also supported by the explicit solutions presented in Section~3. Indeed, for some values of the parameters there exist sphalerons --- the solutions with $\omega=0$ and $E_{ren}\neq 0$ that are always classically unstable \cite{Derrick:1964ww}. Hence we expect that at least the solutions, whose values of $\omega$ are close to $0$, are also classically unstable.

The classical instability of Q-holes and Q-bulges is a very important property of these solutions. It is related to the already mentioned problem of the condensate fragmentation. For example, in the model with the potential (\ref{Potential}) there exist classically stable Q-balls \cite{Theodorakis:2000bz}, along with the classically stable condensate. Therefore, one can expect that Q-holes and Q-bulges represent an important intermediate step in the process of condensate fragmentation (i.e., ``stable condensate''$\to$``unstable Q-hole/Q-bulge''$\to$``stable Q-balls''). In this case, Q-holes or Q-bulges may form due to interactions of the scalar condensate with other particles existing in the theory.

It would be nice to examine analytically the classical instability of the Q-holes and Q-bulges, obtained above for the scalar field potential (\ref{Potential}), by considering the linearized theory on top of the background solution. However, contrary to the case of a simpler potential used in \cite{Gulamov:2013ema,Nugaev:2015rna}, the potential (\ref{Potential}) does not allow for such analysis, the reason being the additional term $\sim\sqrt{\phi^{*}\phi}$ that it contains. There still remains a purely numerical way to investigate the classical instability in the theory with the potential (\ref{Potential}). The numerical analysis may also clarify whether or not Q-holes or Q-bulges lead to fragmentation of the scalar condensate into Q-balls. We leave the thorough investigation of these issues for future work.

\section{Conclusion}
In this paper we have presented Q-holes and Q-bulges --- two classes of localized configurations representing dips and rises in the spatially-homogeneous charged time-dependent scalar condensate. The important feature of these configurations is that they can be deformed into the condensate by a finite amount of energy. We expect that inhomogeneities of this type may be crucial for the nonlinear dynamics of the condensate in the Early Universe, in particular, for its fragmentation into Q-balls. We have also found the explicit solutions for Q-holes in the model with a simple piecewise-parabolic potential proposed in \cite{Theodorakis:2000bz}, and examined their properties. It has been shown that the renormalized energy $E_{ren}$ of Q-holes can take both positive, zero and negative values.

In this paper, we did not address in detail the question of quantum stability of Q-holes and Q-bulges. Of course, if ``ordinary'' particles interact with Q-holes and Q-bulges through, say, the combination $\phi^{*}\phi$ (which is time-independent for these solutions and for the scalar condensate), Q-bulges and Q-holes with $E_{ren}>0$ can decay into such particles. Moreover, one may expect that Q-bulges and Q-holes can be created (even spontaneously) in processes involving these particles. So, this case is rather standard.

However, the case of excitations of the scalar field $\phi$ above the condensate, which are supposed to form the corresponding scalar particles, is not so trivial. For ordinary Q-balls one can define a standard vacuum far from the core of the soliton and apply a standard quantization procedure to the perturbations above this vacuum. For the time-dependent scalar condensate, excitations on top of the background have nonstandard dispersion laws like the one in Eq.~(\ref{excitdispl}). Moreover, one can check that even the charge of the excitation with respect to the condensate charge also has a very nonstandard form, and the standard quantization procedure can not be applied to such excitations. It would be interesting to see what should be defined as ``particles'' related to the excitations of the form (\ref{pertback}) on top of the time-dependent background, and what must be the consistent quantization procedure, providing us with the correct definition of energy of the quantum excitations. These questions call for further detailed investigation.

\section*{Acknowledgements}
We would like to thank D.S.~Gorbunov and V.A.~Rubakov for helpful discussions. The work was supported by grant
14-22-00161 of the Russian Science Foundation (Section~2, Subsections~3.2,~3.3) and by grant NSh-7989.2016.2 of the President of Russian Federation (Subsection~3.1, Section~4).

\section*{Appendix A: limits on $\omega$ for Q-holes in $(3+1)$-dimen\-sional space-time}
In Eq.~(\ref{Rqholes3d}), make a substitution $Y=\sqrt{M^2-\omega^2}R$ and rewrite it in the form
\begin{equation}\label{Rqholes3d2}
\frac{Y\left(\coth(Y)+1\right)}{Y+1}=\frac{\epsilon M^{2}}{M^{2}-\omega^2}.
\end{equation}
The l.h.s. of Eq.~(\ref{Rqholes3d2}) is a monotonic function such that
\begin{equation}
1\le\frac{Y\left(\coth(Y)+1\right)}{Y+1}<2.
\end{equation}
Hence
\begin{equation}
1\le\frac{\epsilon M^{2}}{M^{2}-\omega^2}<2.
\end{equation}
The latter inequality leads to Eq.~(\ref{restr3d}).

\end{document}